\documentclass[prb,showpacs,twocolumn,amsmath,amssymb]{revtex4}

\usepackage{graphicx,amssymb,amsmath,dsfont,epsfig}
\usepackage[latin1]{inputenc}

\newcommand{\pin}{\psi^{in}}
\newcommand{\pout}{\psi^{out}}
\newcommand{\pinout}{\psi^{in/out}}

\begin{document}

\title{Atomistic calculation of the thermal conductance of large scale bulk-nanowire junctions}
\author{Ivan Duchemin}
\affiliation{Max Planck Institute for Polymer Research, Ackermannweg 10, 55128 Mainz, Germany}
\email{duchemin@mpip-mainz.mpg.de}
\author{Davide Donadio}
\affiliation{Max Planck Institute for Polymer Research, Ackermannweg 10, 55128 Mainz, Germany}
\email{donadio@mpip-mainz.mpg.de}

\begin{abstract}
We have developed an efficient scalable kernel method for thermal transport in open systems, with which we have computed the thermal conductance of a junction between bulk silicon and silicon nanowires with diameter up to 10 nm. We have devised scaling laws for transmission and reflection spectra, which allow us to predict the thermal resistance of bulk-nanowire interfaces with larger cross sections than those achievable with atomistic simulations. Our results indicate the characteristic size beyond which atomistic systems can be treated accurately by mesoscopic theories.

\end{abstract}
\maketitle

\section{Introduction}
Nanostructures and nanostructured materials offer the possibility to tune heat transport properties over an exceptionally wide range.
For example in carbon based materials it is possible to obtain variations of the thermal transport coefficients over three orders of magnitude: graphene and suspended carbon nanotubes are possibly the most efficient heat conductors~\cite{Yu:2005p3207,Ghosh:2010p3194}, whereas nanotube pellets and graphene nanoribbons with disordered edges are predicted to have thermal insulating properties~\cite{Prasher:2009p686,Li:2010p3224}. Similarly, nanostructuring may turn silicon and SiGe alloys into efficient thermoelectric materials, by significantly reducing the thermal conductivity ($\kappa$) as in the case of nanowires~\cite{Rurali:2010p2736,Boukai:2008p3187,Hochbaum:2008p3165} (SiNW), SiGe nanocomposites~\cite{Dresselhaus:2007p3571}, superlattices~\cite{Pernot:2010p3443}, and nanoporous silicon~\cite{Yu:2010p3477,Tang:2010p3522}.

 Further improvement in designing materials and nano-devices with controlled thermal transport properties stems from a deeper theoretical understanding of phonon transport. Following Landauer and B\"uttiker's works~\cite{Landauer:1970p3569,BUTTIKER:1985p3568}, atomistic Green's function (GF) formalism has become the reference method to study coherent electronic transport~\cite{WINGREEN:1993p3678,Todorov:1996p3470,Brandbyge:2002p3677}. The GF approach has been transferred successfully to compute thermal transport in nanostructures~\cite{Fagas:1999p3575,Ozpineci:2001p3574,Segal:2003p3577,Mingo:2003p110}, and it is the optimal framework to investigate elastic phonon scattering from impurities, defects, disorder or interfaces, i.e. in all those cases where anharmonic phonon-phonon scattering can be deemed of secondary importance~\cite{Savic:2008p86,Markussen:2009p1869}. An atomistic GF method including phonon-phonon scattering has also been developed and applied to small model systems~\cite{Mingo:2006p3573}, however one can in general safely assume elastic scattering when a finite nanoscale system between two reservoirs is considered. This is generally the case for molecular junctions, contacts and grain boundaries. Special care must be taken in testing this assumption when one wants to extrapolate finite size calculations to extended materials, where long wavelength phonons do not get scattered by nanoscale impurities and contribute a significant amount to the total thermal conductivity.
In spite of significant insight achieved in these studies based on the GF formalism, it remains a formidable task to perform atomistic simulations of nanostructures with characteristic sizes of several tens of nanometers, as it would be needed to bridge the gap between theory and experiment.
Because of matrix inversion operations, even the recursive implementation of the GF method, which permits us to deal with systems extending for several micrometers in the direction of heat propagation, imposes severe size limitations in the orthogonal plane. In terms of SiNW, this means that one is limited to diameters that do not exceed few nanometers. Similar limitations hamper the predictive power of approaches based on molecular dynamics, so far restrained to the study of thin wires~\cite{Donadio:2009p1173,Donadio:2010p3067}.

Here we outline a formalism for phonon transport based on the scattering-matrix approach~\cite{Sautet:1988p3469}, which circumvents the
matrix inversion problem by substituting eigenvalue equations with local kernel search and intersections.
After deriving a generalized scattering-matrix approach for phonon propagation, we illustrate the numerically stable and efficiently parallelizable kernel method.
For example, we apply the scattering formalism to compute the contact thermal resistance between bulk silicon and SiNWs with diameters up to 14 nm.

\section{Scattering matrix approach}
The scattering-matrix approach was formulated to solve quantum electronic transmission problems~\cite{Sautet:1988p3469,Ami:2002p3516}, and found its natural application for the simulation of scanning tunneling microscopy images~\cite{Sautet:1991p3473} and of molecular electronic devices via the so called {\sl elastic scattering quantum chemistry} (ESQC) method~\cite{Ami:2002p3516}.
Here we reformulate the theory in terms of phonon transport.
We consider a phonon wave packet, represented by a weight-normalized displacement field  $u$, traveling through an open system made of semi-infinite reservoirs connected by an arbitrary structure (defect). Our goal is to determine the thermal energy exchanged between the reservoirs through the defect in stationary non-equilibrium conditions, i.e. when the reservoirs are kept at different temperatures.

In the harmonic approximation, the equation of motion for the displacement field $u(t)$ is
$\ddot{u}(t)=\mathbf{D}u(t)$, where \(\mathbf{D}\) is the dynamical matrix. The real-valued state $u$ can be decomposed in terms of the complex valued eigenstates \(v(\omega)\) of \(\mathbf{D}\). Given the state \(u(\tau_{0})\) and its eigen-decomposition coefficients $g_{\tau_{0}}(\omega)$, the time propagation of $u$ is:
\begin{equation}
  u(t)=\int \left[g_{\tau_{0}}(\omega)v(\omega)e^{-i\omega(t-\tau_{0})}+cc.\right] d\omega \ .
  \label{eq_eigendecomposition_ph}
\end{equation}
Let \(\mathbf{P}\) be the projector associated with the degrees of freedom of an arbitrary part P of the system. To get the energy exchanged between P and the rest of the system, one can balance the time derivatives of the work from P to the whole system and vice versa, thus obtaining:
\begin{equation}
  \dot{E}_{P}(t)=\langle\dot{u}(t)|\big[\mathbf{P},\mathbf{D}\big]|u(t)\rangle\ .
  \label{eq_e_var}
\end{equation}
The energy of P in stationary conditions (\(E_{P}(\infty)\)) is found by integrating \ref{eq_e_var} to the infinite time limit. Substituting $u$ with its eigen-decomposition in \ref{eq_eigendecomposition_ph} in the integral leads to:
\begin{equation}
  E_{P}(\infty) = -2\pi i \int \hbar\omega\,|g_{\tau_{0}}(\omega)|^{2}\langle v(\omega)|[\mathbf{P},\mathbf{D}]|v(\omega)\rangle d\omega\,.
  \label{infinite_time_limit_2}
\end{equation}
All information concerning the initial state lies in the weights $g_{\tau_{0}}(\omega)$, which can be taken as the statistical distribution of the states $|v\rangle$ when simulating a system at finite temperature.
In the stationary non-equilibrium case, those weights refer to the rate of phonons emitted from the reservoirs (i.e. 1D phonon gas obeying Bose-Einstein statistics):
\begin{equation}
|g_{0}(\omega)|^{2}=\frac{1}{2\pi}\frac{1}{e^{\hbar\omega/kT}-1}=\frac{1}{2\pi}f(\omega,T),
\label{eq_weight}
\end{equation}
where $f(\omega,T)$ is the Bose-Einstein distribution function at the reservoir temperature $T$.
In order to evaluate \ref{infinite_time_limit_2}, the eigensolutions \(|v(\omega)\rangle\) of the open system have to be expressed in terms of a convenient basis made of a single phonon mode $|\psi^{in}_{i\in A}(\omega)\rangle$ coming from a reservoir \(A\) into the defect, and the set of phonon modes $\psi_{j}^{out}(\omega)$ coming out of the defect toward the reservoirs:
\begin{equation}
|v_{i}(\omega)\rangle=|\psi_{i}^{in}(\omega)\rangle + \sum_{j} S_{ji}(\omega)|\psi_{j}^{out}(\omega)\rangle\ + |v_{i}^{def}(\omega)\rangle ,
\label{basis1}
\end{equation}
where both defect displacements and reservoir surface states at the interfaces are included in $|v^{def}_{i}(\omega)\rangle$.
The scattering tensor $\mathbf{S}(\omega)$ maps the incoming phonons $|\psi^{in}_{i}(\omega)\rangle$ onto the outgoing phonons  $|\psi^{out}_{j}(\omega)\rangle$.
As the energy carried by any incoming or outgoing phonon with frequency $\omega$ is quantized as $\hbar \omega$, \ref{infinite_time_limit_2} provides the following normalization and orthogonality conditions:
\begin{equation}
\begin{split}
& \phantom{\Big[} \langle \psi_{i\in A}^{in}(\omega)|[\mathbf{P}_{A},\mathbf{D}]|\psi_{j\in A}^{in}(\omega)\rangle =-\frac{i\hbar}{2\pi}\cdot \delta_{ij}\\
& \phantom{\Big[} \langle \psi_{i\in A}^{out}(\omega)|[\mathbf{P}_{A},\mathbf{D}]|\psi_{j\in A}^{out}(\omega)\rangle =\phantom{-}\frac{i\hbar}{2\pi}\cdot \delta_{ij}\\
& \phantom{\Big[} \langle \psi_{i\in A}^{in}(\omega)|[\mathbf{P}_{A},\mathbf{D}]|\psi_{j\in A}^{out}(\omega)\rangle = \phantom{-}0 ,
\end{split}
\label{phonon_norm}
\end{equation}
where $\mathbf{P}_{A}$ denotes the projector on reservoir $A$.
Combining the stationary non-equilibrium weights of \ref{eq_weight} with \ref{infinite_time_limit_2}, and observing the conditions of \ref{phonon_norm}, one obtains the stationary energy transfer between reservoirs $A$ and $B$:
\begin{equation}
  \Phi_{A\to B} = \!\int\frac{\hbar\omega}{2\pi}\sum_{i\in A}\sum_{j\in B}|S_{ij}(\omega)|^{2}
  \left[f(\omega,T_{A})-f(\omega,T_{B})\right]d\omega .
  \label{infinite_time_limit_scattering}
\end{equation}
Once \(\mathbf{S}(\omega)\) is obtained by computing the eigenstates \(|v(\omega)\rangle\), the energy flux between two reservoirs $A,B$ is determined using the transmission coefficient $\mathcal{T}_{AB}(\omega)=\sum_{i\in A}\sum_{j\in B}|S_{ij}(\omega))|^{2}$. The corresponding thermal conductance is given by the Landauer formula as the limit of \ref{infinite_time_limit_scattering} when $T_{A}\to T_{B}$:
\begin{equation}
  \sigma_{AB}(T) = \int\frac{\hbar\omega}{2\pi} \mathcal{T}_{AB}(\omega) 
  \dot{f}(\omega,T) d\omega
  \label{landauerformula}
\end{equation}

\section{Scalable implementation}
\begin{figure}[h]
\begin{center}
{\includegraphics[width=55mm]{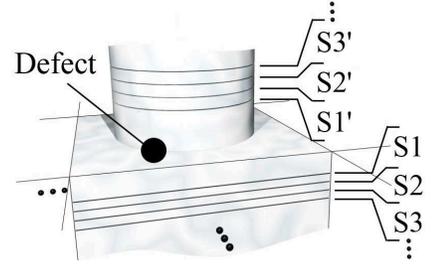}}
\end{center}
\caption{Silicon nanowire/bulk interconnect partitioning. The central defect is generally further subdivided in order to speed up the computation. The bulk and wire reservoirs are organized as a pile of periodic slices (S and S', respectively) indexed starting from the contact areas.}
\label{FigScheme}
\end{figure}
The first step to compute the eigenstates in \ref{basis1} is to rewrite the eigenvalue problem as a null-space search problem:
\begin{equation}
  \mathbf{D}v=\omega^{2}v\quad\Leftrightarrow\quad v\in\ker\{\mathbf{D}-\omega^{2}\}
  \label{master}
\end{equation}
It is then possible to consider a partition \(\mathcal{P}=\{\mathbf{P}_{i}\}\) of the system , and to solve locally the auxiliary equations
\begin{equation}
  v\in\ker\{\mathbf{P}_{i}(\mathbf{D}-\omega^{2})\} .
  \label{auxiliary}
\end{equation}
The solutions of the eigenvalue problem are given by the intersection of the resulting invariant subspaces:
\begin{equation}
  \mathbf{D}v=\omega^{2}v\quad\Leftrightarrow\quad v\in\bigcap_{\mathcal{P}}\ker\{\mathbf{P}_{i}(\mathbf{D}-\omega^{2})\ .
  \label{master_2}
\end{equation}
Typically, the partition is a set of projectors on each reservoir, completed by a set of projectors on the defect. The auxiliary equations (\ref{auxiliary}) for the defect are solved with a QR decomposition method, which is stable and numerically efficient.

The reservoirs are treated in a separate way: as in the propagator method~\cite{Sautet:1988p3469}, every reservoir is partitioned into periodic slices $S_i$ of dimension \(n\) such that only nearest neighboring slices interact (see \ref{FigScheme}). However, instead of formulating a spatial propagator, we first compute the \(2n\) non-trivial solutions of \ref{auxiliary} for the second slice \(S2\) of the reservoir:
\begin{equation}
\mathbf{P}_{S2}(\mathbf{D}-\omega^{2})\cdot
\left(\begin{array}{l}
v^{i}_{S1}\\
v^{i}_{S2}\\
v^{i}_{S3}\\
\end{array}\right)
=0 \ .
\end{equation}
The periodic solutions are then reconstructed  by solving the \(2n\times2n\) generalized eigen-problem:
\begin{equation}
\alpha
\left[\begin{array}{c}
v^{1}_{S1} \dots v^{2n}_{S1}\\
v^{1}_{S2} \dots v^{2n}_{S2}\\
\end{array}\right]\!\cdot\!
\left[\begin{array}{c}
c_{1}\\
\vdots\\
c_{2n}\\
\end{array}\right]
=\beta
\left[\begin{array}{c}
v^{1}_{S2} \dots v^{2n}_{S2}\\
v^{1}_{S3} \dots v^{2n}_{S3}\\
\end{array}\right]\!\cdot\!
\left[\begin{array}{c}
c_{1}\\
\vdots\\
c_{2n}\\
\end{array}\right] .
\end{equation}
As every slice of the reservoir except \(S1\) is equivalent to \(S2\), the periodic solutions hold for the entire reservoir, except for \(S1\) which is treated explicitly as part of the defect. The intersection of the periodic solutions leads to phonon modes \(|\pinout(\omega)\rangle\) (\(|\alpha/\beta|\)=1), and surfaces states (\(|\alpha/\beta|\neq1\)).
After intersecting the reservoir and the defect solutions, we extract a basis \(\{\widetilde{v}_{i}\}\) spanning only surface states localized at the defect interface (i.e. with \(|\alpha/\beta|<1\)):
\begin{equation}
\widetilde{v}_{i}=
\sum_{j}\Big( \Lambda_{ji}|\pin_{j}(\omega)\rangle+\Gamma_{ji}|\pout_{j}(\omega)\rangle\Big)+|\widetilde{v}^{def}_{i}(\omega)\rangle .
\end{equation}
The scattering tensor is then computed by applying the \(\Lambda^{-1}\) transform to the \(\{\widetilde{v}_{i}\}\) set, providing the set of eigenstates \(\{v_{i}\}\) defined in \ref{basis1}, so that:
$\mathbf{S}(\omega)=\Gamma\cdot\Lambda^{-1}$.
In the presence of short range interactions, parts can be defined to be as small as the interaction range, so that only neighboring parts interact. Such an implementation allows efficient parallelization, in the same fashion as domain decomposition in molecular dynamics codes. The auxiliary equations are solved locally in parallel before their intersection according to a binomial tree. Furthermore, a reciprocal space sampling technique allows an efficient treatment of the periodic reservoirs solutions. Within this framework, the main limitation of the approach is the treatment of non-periodic 1D reservoirs, which requires the full diagonalisation of a matrix, which scales as the square of the surface of the contact.

\section{Results and discussion}
\paragraph{Thermal conductance of bulk silicon/silicon nanowire contacts.} We apply the scattering matrix approach to compute the contact thermal resistance of a bulk-SiNW interface. 
Interface resistance plays an essential role in determining the thermal transport performance of nanostructured materials and nanoscale devices. In addition, evaluating the thermoelectric performances of nanostructures such as SiNW, it is indispensable to be able to resolve the contact thermal resistance from the intrinsic resistance.
A few special cases, such as grain boundaries in silicon, crystalline/amorphous interfaces and silicon/germanium junctions, have previously been addressed using molecular dynamics and real-space Kubo-Greenwood formalism~\cite{Bodapati:2006p3523,Landry:2009p3275}.
An often overlooked yet omnipresent case where contact resistance is essential is the junction between nanostructures and reservoirs. A simplified model, based on lattice dynamics calculations of bulk silicon and of SiNWs with different diameters, predicts that the contact resistance is dominant over the intrinsic resistance of the ideal nanowire~\cite{Chalopin:2008p3483,Chang:2005p3580}.
Coherent contacts between crystals and nanowires with diameter as small as $\sim 20$ nm can be actually realized by etching nanowires directly out of the bulk precursor~\cite{Heron:2009p1725,Hippalgaonkar:2010p3572}.
We model the interatomic interactions between silicon atoms by means of the short-range empirical forcefield after Tersoff~\cite{JTersoff1989}. Crystalline SiNWs with diameters between 2 and 14 nm are considered. The wires are grown in the (100) crystallographic direction, have a nearly circular cross section and are coherently connected to the bulk reservoir. The surface is reconstructed in order to minimize the number of dangling bonds~\cite{Vo:2006p681}.
\begin{figure}[h!]
\begin{center}
\includegraphics[width=75mm]{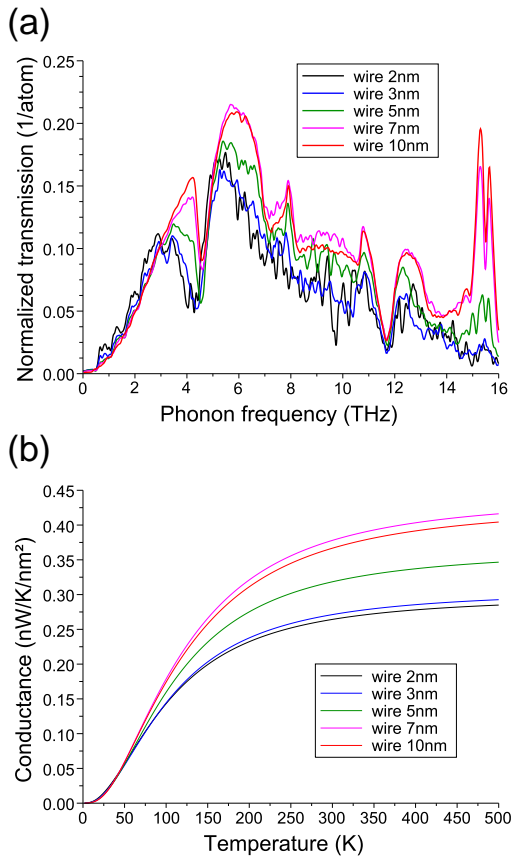}
\end{center}
\caption{(color) Transmission spectra (a) and thermal conductance as a function of the temperature (b) for a set of bulk-nanowire contacts. Both data sets are normalized with respect to the interface area expressed either in nm (conductance) or in number of atoms (transmission).}
\label{Fig:transmission}
\end{figure}

Transmission spectra are displayed in \ref{Fig:transmission}(a) along with the interface conductance $\sigma$ (b) obtained by integrating the transmission coefficient over the whole frequency spectrum according to \ref{landauerformula}. The data sets are normalized according to the interface area, as one would reasonably expect the conductance of a junction to scale with its cross section area. In fact such normalization makes curves comparable, but not overlapping. Normalized transmission spectra and $\sigma(T)$ curves overlap for wires of 7 and 10 nm diameter (red and pink curves in \ref{Fig:transmission}). 
Whereas heat transport in thicker wires can be treated within a mesoscopic approach~\cite{Angelescu:1998p3666,Prasher:2006}, below this threshold, one has to consider explicitly the atomistic details of the interface to obtain an accurate estimate of the contact conductance. As the construction  of the bulk-wire interface is ideal at the atomic scale, our calculations provide an upper limit to the contact conductance.
In the low temperature regime ($T < 50$ K) the interface area normalized contact conductances collapse to a single curve and display a temperature dependence of $T^3$. This trend was formerly predicted analytically~\cite{Chang:2005p3580} and confirmed in experiments~\cite{Heron:2009p1725}, where it was shown that deviations from the $T^3$ behavior stem from specific features of the SiNW, such as surface roughness, the effects of which add up in series to the contact conductance.
On the other hand we observe that the reflection spectrum (not shown), scales with the linear dimension of the contact interface, which seems to indicate that back scattering of phonons mostly happens at the perimeter of the junction. Normalized reflection spectra nearly coincide for wires with a diameter larger than 7 nm.

\begin{figure}[t]
\begin{center}
{\includegraphics[width=95mm]{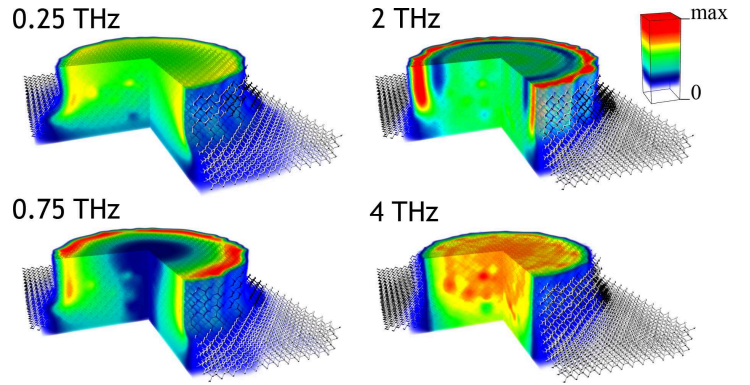}}
\end{center}
\caption{(color) Volumetric representation of the norm of the energy flux at the interface of a 10 nm thick silicon nanowire, corresponding to channels with frequency of 0.25, 0.75, 2 and 4 THz. In the 0.75 and 2 THz case, thermal transport mainly occurs in a thin sub-surface layer (red color area).}
\label{Fig:fluxes}
\end{figure}

\paragraph{Representation of the heat flux.} An advantage of the present implementation of the scattering matrix method is that it provides a real-space representation of the energy flux at any given frequency. This allows visualization of the parts of the system that primarily transmit or reflect thermal energy. An example is shown in \ref{Fig:fluxes}, where the norm of the heat flux across a bulk-10nm SiNW interface is represented. Phonon branches at 0.25, 0.75, 2 and 4 THz are considered. The spacial features of heat transport at different frequencies are clearly different: whereas at the lowest frequency (0.25 THz) thermal energy is mainly transmitted through the central bulk-like part of the wire, at higher frequencies (0.75 and 2 THz) thermal energy is transferred through a surface layer. Beyond 4 THz heat is transferred through the center of the wire. We note that phonons with frequency between $\sim 1$ and $\sim 4$ THz, which are the majority heat carriers in crystalline Si at room temperature, transfer energy preferably through a sub-surface layer.
Therefore our results indicate the reason why thermal conductivity of SiNW is so sensitive to surface modifications, such as disorder or presence of interfaces~\cite{Hochbaum:2008p3165,Hu:2010p3581}.

\begin{figure}[t]
\begin{center}
{\includegraphics[width=75mm]{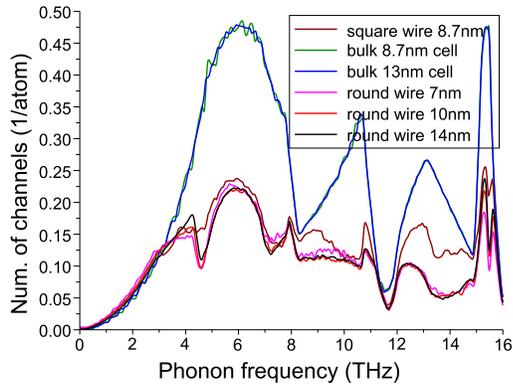}}\\
\end{center}
\caption{(color) Number of transmission channels for a set of bulk-nanowire contacts of different diameter and shape. The data are normalized with respect to the interface area expressed in number of atoms. Data are compared to the number of channels in a three-dimensional periodic bulk to highlight the effect of dimensionality reduction. All the nanowires considered here are larger than the 7 nm diameter threshold.}
\label{Fig:channels}
\end{figure}

%
\paragraph{Dimensionality and shape effects.} In order to probe the effects of shape and dimensionality reduction on the contact conductance we compare the number of phonon channels (corresponding to the density of states) over the whole frequency spectrum, in contacts made of crystalline bulk silicon and either wires with a circular section or square rods. We only consider SiNW larger than the threshold size of 7 nm, identified as the onset for a mesoscopic theory of thermal transport. The calculations have been performed for SiNW with diameters up to 14 nm. The data are conveniently normalized with respect to the contact surface area and are compared to the number of channels in three-dimensional periodic bulk. To verify size convergence we consider two bulk samples with cubic supercell of 8.7 and 13 nm, respectively (\ref{Fig:channels}).
Our data show that for SiNWs larger than 7 nm, the number of channels per atom at a given frequency does not depend on the diameter. The number of channels at low frequency ($<3$ THz) for the contacts is the same as in the crystalline bulk, but it deviates significantly from the bulk at larger frequencies. This means that even in contact interfaces with very large wires, one cannot expect to recover bulk-like thermal conductance. It also indicates that dimensionality reduction has a profound effect on the limit density of states as well. Such a limit depends also on the shape of the SiNW, but to a minor extent. The spectrum of square shaped nanorods differs from that of circular ones in the medium-to-high frequency range, but it retains similar features as cylindric wires and does not seem to approach the 3D bulk limit either. 

\section{Conclusions}
We have developed an efficient method based on the scattering matrix approach to compute the thermal conductance in an open system. Our derivation leads to an expression of the energy flux between two reservoirs across a defect region, equivalent to the one derived in Refs.~\cite{Fagas:1999p3575,Ozpineci:2001p3574,Segal:2003p3577}. However, our implementation differs from the traditional Green's function approach, as it circumvents the bottleneck imposed by the inversion of large matrices, and allows real size devices to be simulated at the atomistic level. We have used this approach to compute the contact thermal conductance of ideal junctions between  bulk silicon and silicon nanowires of different diameters. Our results show that beyond a threshold diameter of 7 nm phonon transmission, reflection and thermal conductance obey simple scaling laws, whereas deviations are observed for thinner wires. Our approach also provides a direct space visualization of frequency dependent heat flux, which yields valuable insight into the spatial features of heat conduction in nanoscale devices.

\begin{acknowledgements}
Calculations were performed on the IBM Power6 system at the Rechenzentrum Garching.
I.D. acknowledges support from the Multiscale Materials Modeling Initiative of the Max Planck Society.
The authors thank K. Kremer and C. Joachim for useful suggestions, and L.F. Pereira and L. Pavka for critical reading of the manuscript.
\end{acknowledgements}

\end{document}